\documentclass[psfig,12pt]{article}
\usepackage{color}
\usepackage{psfig}

\begin{document}

\newcommand{\simless}{\mathbin{\lower 3pt\hbox
     {$\rlap{\raise 5pt\hbox{$\char'074$}}\mathchar"7218$}}}

\bigskip

\centerline{\Large\bf Formation of the Widest Binaries}
\vspace{0.4cm}
\centerline{\Large\bf from Dynamical Unfolding of Triple Systems}

\bigskip

\centerline{\large Bo Reipurth$^{1}$ 
        and 
        Seppo Mikkola$^{2}$
}


\begin{center}
{1: Institute for Astronomy, Univ. of Hawaii at Manoa, 
          640 N. Aohoku Place, HI 96720, USA}\\
{reipurth@ifa.hawaii.edu}\\
{2: Tuorla Observatory, University of Turku, V\"ais\"al\"antie
    20, Piikki\"o, Finland}\\
{Seppo.Mikkola@utu.fi}
\end{center}

{\bf The formation of very wide binaries$^{1-3}$, such as the $\alpha$
  Cen system with Proxima (also known as $\alpha$~Centauri~C)
  separated from $\alpha$~Centauri (which itself is a close binary
  A/B) by 15000~astronomical units$^{4}$ (1~AU is the distance from
  Earth to the Sun), challenges current theories of star formation,
  because their separation can exceed the typical size of a collapsing
  cloud core. Various hypotheses have been proposed to overcome this
  problem, including the suggestion that ultra-wide binaries result
  from the dissolution of a star cluster -- when a cluster star
  gravitationally captures another, distant, cluster star$^{5-7}$.
  Recent observations have shown that very wide binaries are
  frequently members of triple systems$^{8,9}$ and that close binaries
  often have a distant third companion$^{10-12}$.  Here we report
  Nbody simulations of the dynamical evolution of newborn triple
  systems still embedded in their nascent cloud cores that match
  observations of very wide systems$^{13-15}$.  We find that although
  the triple systems are born very compact -- and therefore initially
  are more protected against disruption by passing stars$^{16,17}$ --
  they can develop extreme hierarchical architectures on timescales of
  millions of years as one component is dynamically scattered into a
  very distant orbit.  The energy of ejection comes from shrinking the
  orbits of the other two stars, often making them look from a
  distance like a single star.  Such loosely bound triple systems will
  therefore appear to be very wide binaries.}\\

Evidence is building that stars often, and possibly always, are formed
in small multiple systems$^{18,19}$. Dynamical interactions between
members of such systems lead to close triple encounters in which
energy and momentum is exchanged, typically causing the disintegration
of the triple system, with the escape of a single component (most
frequently the lowest mass member) and the formation of a stable
binary$^{20-22}$. A bound triple with a hierarchical architecture may
also result, but only if it forms in the presence of a gravitational
potential can such a triple system achieve long-term stability$^{23}$.
However, this is frequently fulfilled for newborn triple systems,
since break-up typically occurs in the protostellar phase, when the
newborn stars are still deeply embedded in their nascent cloud
cores$^{24}$.

We have used an advanced Nbody code to run 180,218 simulations of a
newborn triple system placed in a gravitational potential$^{23}$ (for
technical details see Supplementary Information).
Figure 1a shows results for the 13,727 stable hierarchical systems that
are formed in the 180,218 simulations. The blue dots mark the semimajor
axes of the inner binaries, while the red dots indicate the semimajor
axes of the outer components relative to the center-of-mass of the
inner binary. The distant components have semimajor axes that span
from hundreds of AU to several thousands of AU, with a small but not
negligible number of cases reaching several tens of thousands of AU
and beyond.  

We run the simulations for 100 million years, and classify
the outcome at 1, 10 and 100~Myr into stable triples, unstable
triples, and disrupted systems. If the outer orbit is hyperbolic, then
the system is disrupted. If not, then the system is (at least
temporarily) bound, and a stability criterion is applied$^{25}$. If
the system passes this test, it is classified as stable. If not, it is
still bound but internally unstable, and will sooner or later disrupt.
Figure~1b shows the semimajor axis distribution for 56,957 bound but
unstable triple systems.

The number of systems in each category is merely an estimate, since a
rigorous theory of stability is not mathematically possible, because
the three-body problem is non-integrable. Hence we do see that the
number of stable triples at 1, 10, and 100~Myr is not completely
constant, but declines a few percent with time. The number of unstable
triples, on the other hand, dramatically declines with age, 
by a factor of 3 or more from 1 to 100~Myr (Fig.~2a). At 1~Myr, 39\%
of systems are bound (stable and unstable), at 10~Myr this number has
decreased to 18\%, and at 100~Myr it is 12\%. This number will
continue to decrease until only the stable triple systems remain,
which is $\sim$7\%.  This last number compares well with the 8\% of
triple systems (mean of all spectral types) observed in the
field$^{26}$, suggesting that most star forming events must start as
triple systems.

The reason for the difference in distribution of semimajor axes
between stable and unstable triple systems (Figs.~1a,b) lies in their
orbital parameters. Figure~2b shows the separation distribution
function of stable outer (red) and inner (blue) and unstable outer
(green) and inner (aqua) systems at an age of 1 Myr.  Figure~2c shows
the distribution of eccentricities among the bound triple systems.
The primary reason that some systems are stable and others unstable is
that the stable systems are well separated even at periastron, when
the three bodies have their closest approach. In contrast, the
unstable systems are much more likely to suffer perturbations at
closest approach, ultimately leading to their disruption.  This is
reflected in the eccentricity of the systems, see the figure caption
for details.

There are three main parameters that control the stability of a
triple system: the semimajor axis, the eccentricity, and the ratio of
periastron distance of the outer binary to apastron distance of the
inner binary. If the outer periastron distance becomes smaller than
roughly 5--10 times the inner apastron distance, then the system will
eventually break up. Systems with close inner binaries thus have a
larger chance of achieving stability. For the outer binary, the
relation between eccentricity and semimajor axis is shown in
Figure~2d. Wide binaries (1,000 -- 10,000~AU) are found with all
eccentricities except the very smallest, although with a clear
preference for larger eccentricities. For the very wide binaries
($>$10,000~AU), in contrast, the eccentricities $e$ tend to be extreme
(red dots in Figure~2d). The reason they can survive is that their
periastron distance $a(1-e)$ is also large, thanks to their semimajor
axes $a$ also being extremely large, although a few are stable even
with moderate eccentricity, presumably because they have unusually
small inner binaries.

Wide stable and unstable triple systems differ in one
  important respect.  Figure~3a,b show how total binary mass relates
  to mass of the distant third body for systems wider than 1,000~AU at
  ages of 1~Myr and 100~Myr. A large population exists at 1~Myr with
  members that are either all three approximately of the same mass, or
  with the distant third member being of very low mass. This
  population, however, is largely unstable (green), and so has mostly
  disappeared by 100~Myr, except for the very low-mass systems.  The
  reason that systems with a dominant binary and a light single are
  more unstable than the opposite configuration is likely that massive
  binaries more easily can alter the orbit of the third body near
  periastron, eventually leading to disruption. Stable triple systems
  (red) are much more uniformly distributed across Figure~3, although
  with a preference for members of very low-mass systems to be
  approximately of the same mass, and a slight preference for systems
  with dominant singles rather than dominant binaries. Such
  time-dependent properties of wide triple systems may be a dynamic
  signature of the triple decay mechanism.

Our simulations do not take into account that there
  may be further orbital evolution of the inner binary when the decay
  from non-hierarchical to hierarchical configuration occurs during
  the protostellar phase (as it does for more than 50\% of
  simulations$^{23}$). In that case, viscous evolution will cause
  further inspiraling, leading to the formation of spectroscopic
  binaries. Gas induced orbital decay can ultimately lead to the
  merger of the binary components in a non-negligible number of
  cases$^{27}$. It follows that although wide binaries formed through
  triple decay initially consist of three stars, they may during the
  pre-main sequence phase evolve into a true wide binary containing
  only two stars.

  The results presented here refer to the {\em birth population} of
  binaries, at an age of 1~Myr. The orbital parameters of a triple
  system are established at the moment when a stable hierarchical
  triple is formed.  But since the birth configuration of a triple
  system is compact, it will take half an orbital period of the outer
  component before the triple system reaches its first apastron
  passage and attains its maximum extent $a(1+e)$. We call this the
  initial {\em unfolding time} of the newborn triple system.  Many
  wide systems have not unfolded fully at 1~Myr, and the most extreme
  wide systems will take tens to hundreds of million years to unfold,
  and they are thus more protected against disruption by passing
  stars$^{16,17}$ than if triple systems were born with such enormous
  separations in their crowded natal environments (Figure~3c). See
  Supplementary Information for more details.

Non-hierarchical systems that have broken up shortly after birth lead
to a close binary and a detached single star that is moving 
away from the binary. Those with small velocity differences, or with
motions mainly along the line of sight, will be observed to linger for
a while in the vicinity of each other, mimicking a bound binary.  It
is of interest to determine what is the fraction of stable bound,
unstable bound, and unbound triple systems that exist at different
{\em projected} separations at different ages. Figure~4 shows the
number and fraction of systems in each category for 1, 10, and
100~Myr.  As a specific example, the recently discovered very wide
($\sim$12--40~kAU) triple systems found in 7-10~Myr old
associations$^{28,29}$ have only a $\sim$20\% chance of being
long-lived systems, a $\sim$20\% chance of already being disrupted
systems, but a $\sim$60\% chance that they are unstable still bound
systems.

At 100~Myr, 8.5\% of the 180218 simulations have led to bound (stable
and unstable) systems with semimajor axes $a$ between 1,000 and
10,000~AU, and 2.1\% are bound with $a$ $>$ 10,000~AU. That is, more
than 10\% of the birth population of triple systems end up as wide or
very wide, in excellent agreement with observations$^{30}$.  The
simulations also broadly reproduce the observed distribution of
projected separations$^{13-15}$, see Supplementary Information for
details.  The present N-body simulations have thus demonstrated that
the widest binaries known can arise naturally as a consequence of
three-body dynamics shortly after birth.  The subsequent, lengthy
unfolding of the widest systems offers increased protection against
external disruption by other young stars, and allows such wide systems
to be born in both stellar associations as well as the outskirts of
dense clusters.

\clearpage

[1] Bahcall, J.N. \& Soneira, R.M. 
The distribution of stars to V = 16th magnitude near the north galactic pole - Normalization, clustering properties, and counts in various bands.
{\em Astrophys. J.} {\bf 246}, 122-135 (1981).

[2] Sesar, B., Ivezic, Z., \& Juric, M.
Candidate Disk Wide Binaries in the Sloan Digital Sky Survey.
{\em Astrophys. J.} {\bf 689}, 1244-1273 (2008).

[3] Dhital, S., West, A.A., Stassun, K.G., \& Bochanski, J.J.
Sloan Low-mass Wide Pairs of Kinematically Equivalent Stars (SLoWPoKES): A Catalog of Very Wide, Low-mass Pairs.
{\em Astron. J.} {\bf 139}, 2566-2586  (2010).

[4] Wertheimer, J.G. \& Laughlin, G. 
Are Proxima and $\alpha$ Centauri gravitationally bound?
{\em Astron. J.} {\bf 132}, 1995-1997 (2006).

[5] Kouwenhoven, M.B.N. et al. 
The formation of very wide binaries during the star cluster dissolution phase.
{\em Mon. Not. R. Astron. Soc.} {\bf 404}, 1835-1848  (2010).

[6] Moeckel, N. \& Clarke, C.J. 
The formation of permanent soft binaries in dispersing clusters.
{\em Mon. Not. R. Astron. Soc.} 415, 1179-1187 (2011).

[7] Moeckel, N. \& Bate, M.R. 
On the evolution of a star cluster and its multiple stellar systems following gas dispersal.
{\em Mon. Not. R. Astron. Soc.} {\bf 404}, 721-737 (2010).

[8] Law, N.M., Dhital, S., Kraus, A., Stassun, K.G., \& West, A.A.
The high-order multiplicity of unusually wide M dawrf binaries: 
Eleven new triple and quadruple systems.
{\em Astrophys. J.} {\bf 720}, 1727-1737 (2010).

[9] Faherty, J.M. et al.  
The brown dwarf kinematics project. II.  Details on nine wide common
proper motion very low mass companions to nearby stars.
{\em Astron. J.} {\bf 139}, 176-194 (2010).

[10] Tokovinin, A., Thomas, S., Sterzik, M., \& Udry, S. 
Tertiary companions to close spectroscopic binaries.
{\em Astron. Astrophys.} {\bf 450}, 681-693 (2006).

[11] Tokovinin, A. \& Smekhov, M.G. 
Statistics of spectroscopic sub-systems in visual multiple stars.
{\em Astron. Astrophys.} {\bf 382}, 118-123 (2002).

[12] Allen, P.R., Burgasser, A.J., Faherty, J.K., \& Kirkpatrick, J.D.
Low-mass tertiary companions to spectroscopic binaries. I. Common proper motion
survey for wide companions using 2MASS.
{\em Astron. J.} {\bf 144}:62 (2012).

[13] Chanam\'e, J. \& Gould, A. 
Disk and Halo Wide Binaries from the Revised Luyten Catalog: 
Probes of Star Formation and MACHO Dark Matter.
{\em Astrophys. J.} {\bf 601}, 289-310 (2004).

[14] L\'epine, S. \& Bongiorno, B.
New Distant Companions to Known Nearby Stars. II. Faint Companions of
Hipparcos Stars and the Frequency of Wide Binary Systems.
{\em Astron. J.} {\bf 133}, 889-905 (2007).

[15] Tokovinin, A. \& L\'epine, S.
Wide companions to HIPPARCOS stars within 67 pc of the Sun.
{\em Astron. J.} {\bf 144}:102   (2012).

[16] Retterer, J.M. \& King, I.R. 
Wide binaries in the solar neighborhood.
{\em Astrophys. J.} {\bf 254}, 214-220 (1982).

[17] Weinberg, M.D., Shapiro, S.L., \& Wasserman, I. 
The dynamical fate of wide binaries in the solar neighborhood.
{\em Astrophys. J.} {\bf 312}, 367-389 (1987).

[18] Goodwin, S.P., Kroupa, P., Goodman, A., \&
     Burkert, A. 
The Fragmentation of Cores and the Initial Binary Population.
in {\em Protostars and Planets V}, eds. B.
     Reipurth, D. Jewitt, K. Keil, Univ. of Arizona Press, Tucson, 133-147 (2007).

[19] D\^uchene, G., Delgado-Donate, E., Haisch, K.E.,
     Loinard, L. \& Rodr\'\i guez, L.F. 
New Observational Frontiers in the Multiplicity of Young Stars.
 in {\em Protostars and
       Planets V}, eds. B.  Reipurth, D. Jewitt, K. Keil, Univ. of
     Arizona Press, Tucson,  379-394 (2007).

[20] Anosova, J.P. 
Dynamical evolution of triple systems.
{\em Astrophys. Spa. Sci.} {\bf 124}, 217-241 (1986).

[21] Delgado-Donate, E.J., Clarke, C.J., Bate, M.R., \&
  Hodgkin, S.T. 
On the properties of young multiple stars.
{\em Mon. Not. R. Astron. Soc.} {\bf 351}, 617-629 (2004).

[22]  Valtonen, M. \& Mikkola, S. 
The few-body problem in astrophysics.
 {\em Ann. Rev. Astr. Ap.} {\bf 29}, 9-29 (1991).

[23] Reipurth, B., Mikkola, S., Connelley, M., \&
  Valtonen, M.  Orphaned Protostars.  
{\em Astrophys. J.} {\bf 725}, L56-L61 (2010).

[24] Reipurth, B. 
Disintegrating Multiple Systems in Early Stellar Evolution.
{\em Astron. J.} {\bf 120}, 3177-3191 (2000).

[25]  Mardling, R. A. 
Resonance, chaos and stability - Secular evolution of triple systems.
in {\em The Cambridge N-body
    Lectures}, eds. S.J Aarseth, C.A. Tout, R.A. Mardling, 
Springer, Berlin, 59-96 (2008).

[26]  Tokovinin, A. 
Multiple Stars: Designation, Catalogues, Statistics.
in {\em Multiple Stars Across
     the H-R Diagram}, eds. S. Hubrig, M. Petr-Gotzens, A.
     Tokovinin, Springer, Berlin, 38-42 (2008).

[27] Korntreff, C., Kaczmarek, T., \& Pfalzner, S.
Towards the field binary population: influence of orbital decay 
on close binaries.
{\em Astron. Astrophys.} {\bf 543}:A126 (2012).

[28]  Teixeira, R. et al. 
SSSPM J1102-3431 brown dwarf characterization from accurate proper motion and trigonometric parallax.
{\em Astron. Astrophys.} {\bf 489}, 825-827  (2008).

[29] Kastner, J.H. et al. 
2M1155-79 (= T Chamaeleontis B): A Low-mass, Wide-separation Companion to the nearby, "Old" T Tauri Star T Chamaeleontis.
{\em Astrophys. J.} {\bf 747}:L23  (2012).

[30] Longhitano, M. \& Binggeli, B. 
The stellar correlation function from SDSS. A statistical search for wide binary stars.
{\em Astron. Astrophys.} {\bf 509}:A46 (2010).

\vspace{0.3cm}

\noindent
{\bf Supplementary Information} is linked to the online version of the
paper at www.nature.com/nature.

\vspace{0.3cm}

\noindent
{\bf Acknowledgments} We thank two anonymous referees and C.J. Clarke,
M.B.N.  Kouwenhoven, and A. Tokovinin for comments.  BR thanks ESO and
Tuorla Observatory for hospitality during the period when this paper
was written, and Hsin-Fang Chiang and Colin Aspin for providing
additional computer facilities.  This work was supported by the
National Aeronautics and Space Administration through the NASA
Astrobiology Institute under Cooperative Agreement No. NNA09DA77A
issued through the Office of Space Science.  This research has made
use of the SIMBAD database, operated at CDS, Strasbourg, France, and
of NASA's Astrophysics Data System Bibliographic Services.

\vspace{0.3cm}

\noindent
{\bf Author Contributions} BR conceived the idea, carried out the
simulations and data analysis, and wrote the paper. SM developed the
code and wrote the software tools for analysis.

\vspace{0.3cm}

\noindent
{\bf Author Information} Reprints and permissions information is
available at www.nature.com/reprints.  The authors declare no
competing financial interests. Readers are welcome to comment on the
online version of this article at www.nature.com/nature.
Correspondence and requests for materials should be addressed to
reipurth@ifa.hawaii.edu

\clearpage

\begin{figure}[h]\begin{center}\leavevmode
\psfig{file=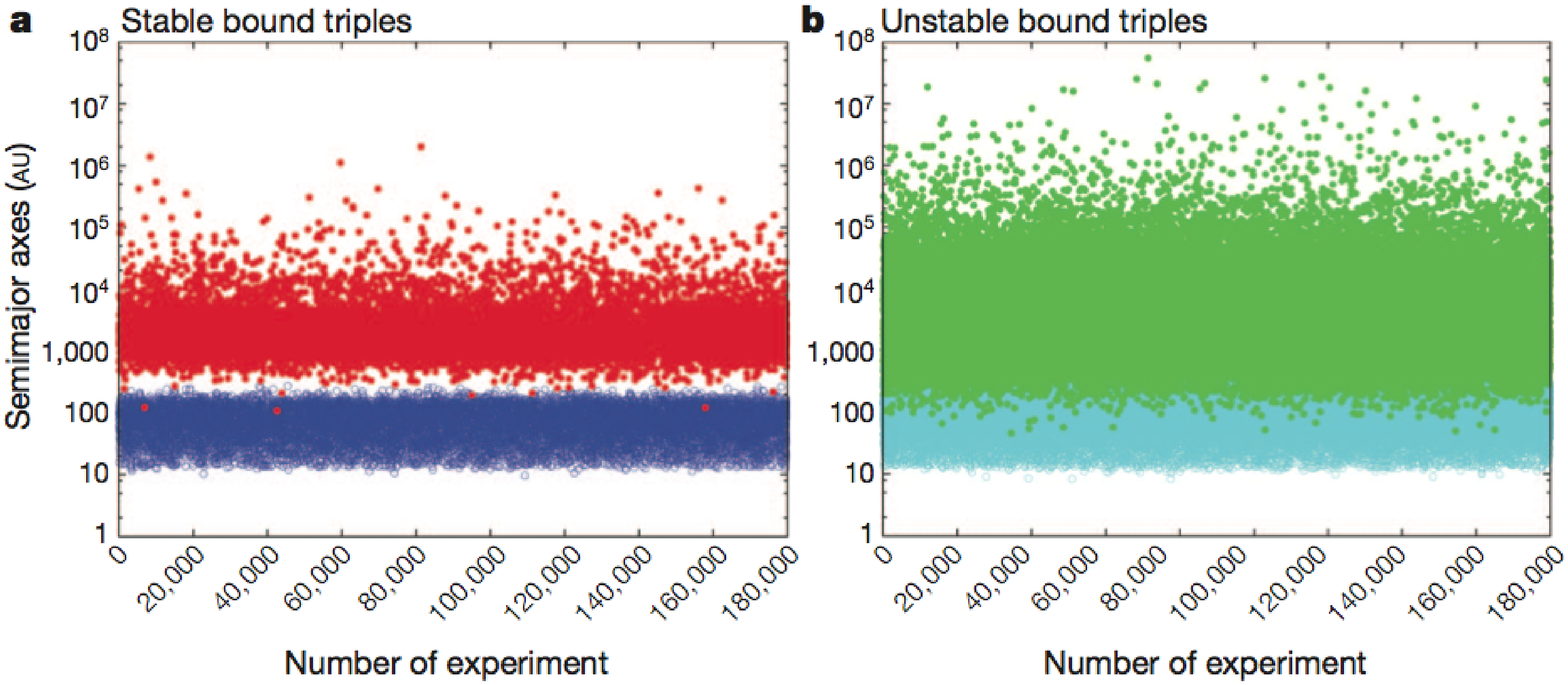,height=7.5cm}
\end{center}
\end{figure}

{\bf Figure 1 $|$ Semimajor axes of stable and unstable bound triple systems.}
(a): The semimajor axes of the outer and inner pairs in 
bound stable triple systems. Filled blue circles are the inner binaries in 
stable hierarchical triple systems, filled red circles are the more distant 
singles in stable hierarchical triple systems. 
(b) The semimajor axes of the outer and inner pairs in bound unstable triple 
systems. Turquoise circles are binaries in bound unstable triple systems, 
and green circles are singles in bound unstable triple systems. For both 
figures the semimajor axes refer to orbital parameters for systems that 
still remain bound at 1 Myr; at later times many of the unstable systems will
have disrupted. For the widest systems, the distant bodies have not yet 
reached their extreme apastron distances.


\clearpage

\begin{figure}[h]\begin{center}\leavevmode
\psfig{file=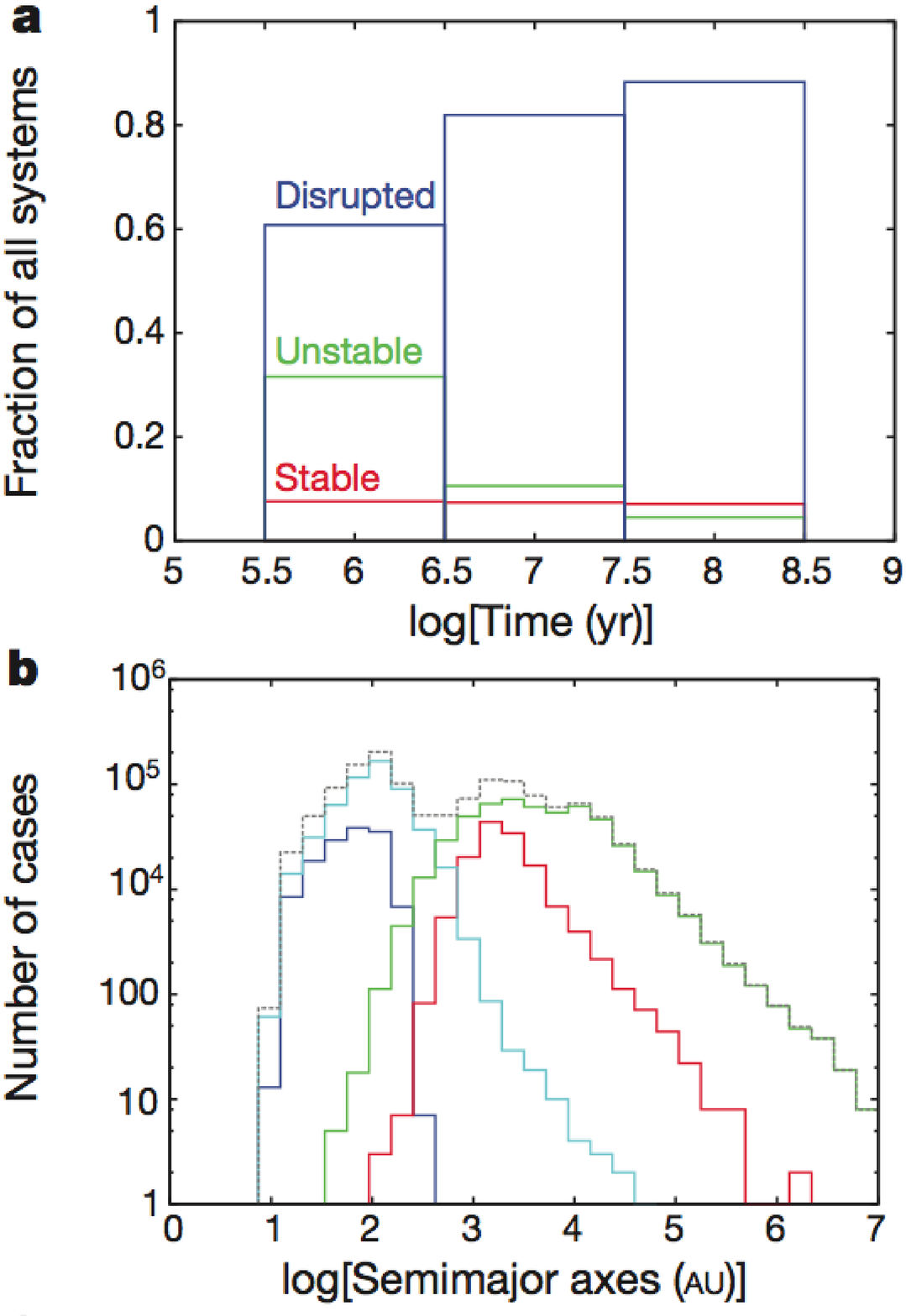,height=10.5cm}
\hspace{1cm}
\psfig{file=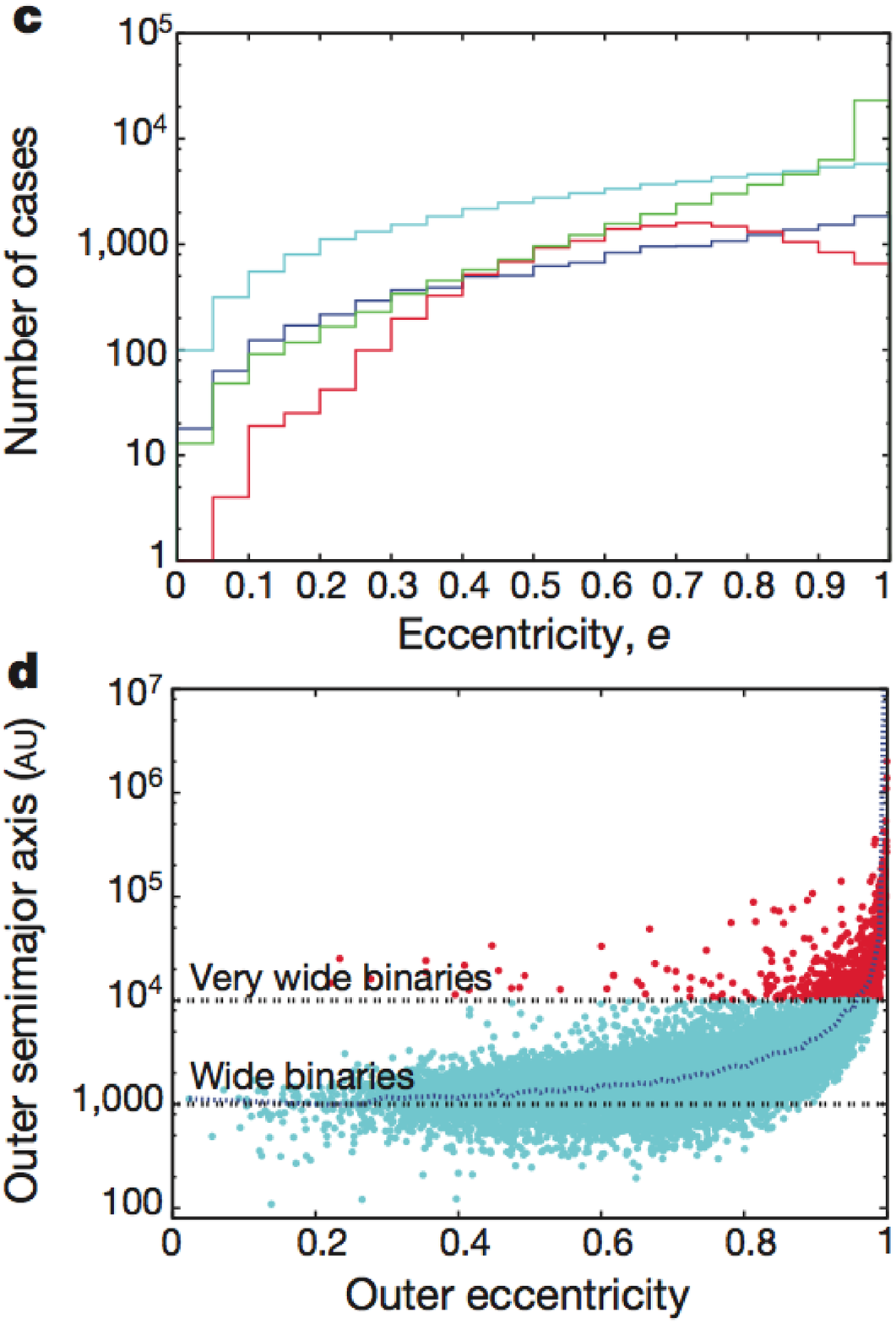,height=10.5cm}
\end{center}
\end{figure}

{\bf Figure 2. $|$ Statistical properties of stable, unstable, and
  disrupted triple systems.}  (a): Histogram with number of stable
hierarchical, unstable hierarchical, and disrupted triple systems at
1, 10 and 100 Myr. The stable systems are essentially constant, while
many of the unstable systems disrupt.  (b): The distribution of
semimajor axes for both inner and outer binaries that are bound at 1
Myr. The color scheme is the same as in Figure~1.  The grey dashed
line shows the sum of all (inner and outer) binaries.  (c): The
distribution of eccentricities for bound inner and outer binaries at
1~Myr. The color scheme is the same as in Figure~1.  It is evident
that highly eccentric systems are common, and the number of triple
systems is a growing function of eccentricity for all but the stable
outer systems, which peak around $e$$\sim$0.7. Systems with very high
eccentricity tend to have smaller periastron distances $a(1-e)$,
leading to the possibility of perturbations which after one or more
close periastron passages eventually lead to breakup.  Hence, the
decline seen at high eccentricities for bound stable systems (red) is
compensated by an increase among the unstable outer systems (green).
The eccentricity distribution for the inner binaries will evolve
significantly if circumstellar material is still present at birth or
due to Kozai cycles.  (d): The distribution of
  semimajor axes of the outer components in triple systems show a very
  strong dependence on the eccentricity. Wide binaries (1,000 $<$ $a$
  $<$ 10,000~AU) get in the mean increasingly wide as the eccentricity
  increases. For very wide binaries ($a$ $>$ 10,000~AU, marked in red)
  this correlation becomes even more pronounced. While very wide
  binaries can be found with modest eccentricities ($e$$\sim$0.3-0.4),
  the majority have eccentricities exceeding 0.9.

\clearpage

\twocolumn

\begin{figure}[t]\begin{center}\leavevmode
\psfig{file=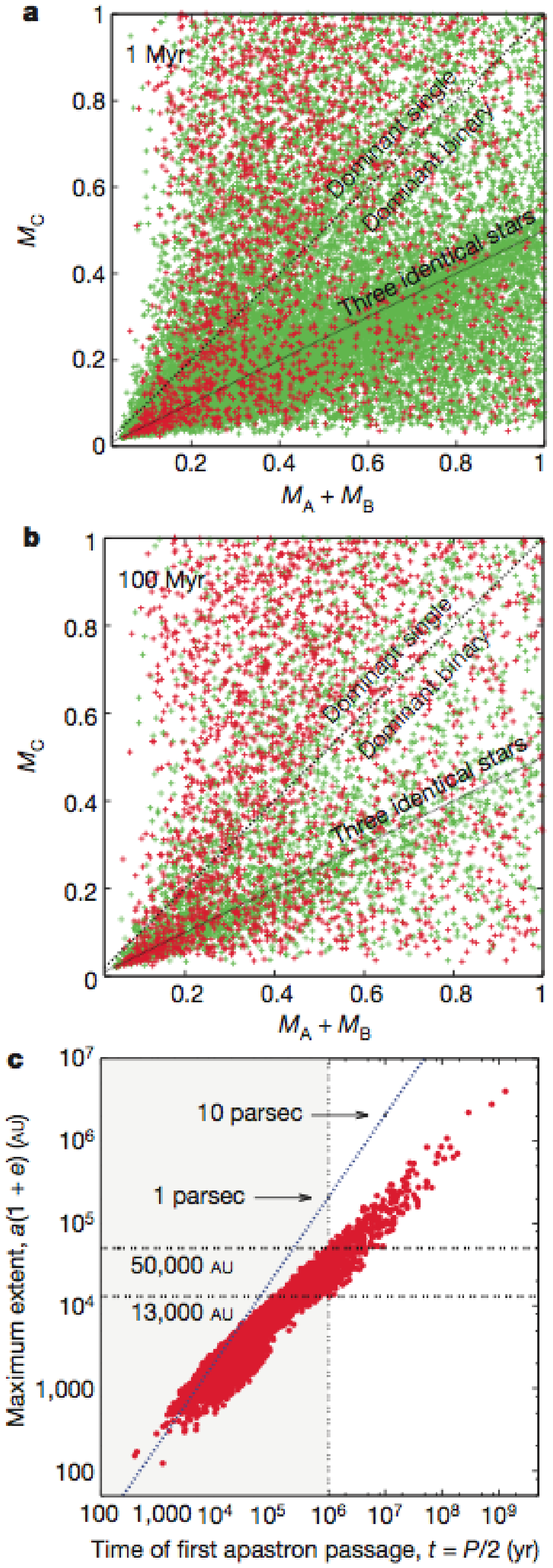,height=22.0cm}
\end{center}
\end{figure}

{\bf Figure 3. $|$ Stable and unstable hierarchical triple systems at
  1 Myr and 100 Myr, and the maximum extent a(1+e) of a triple system
  as a function of time.}  (a) The total mass of the close binary is
plotted against the mass of the distant third body at an age of 1~Myr
for all wide systems with outer semimajor axes exceeding 1,000~AU.
Systems that are classified as unstable are marked green, and systems
that are stable over long timespans are marked red. The figure is
divided into two areas, in one half most of the system mass resides in
the binary, whereas in the other half the single dominates the system.
A line indicates where systems with three identical bodies lie. (b)
The same figure for wide systems at an age of 100~Myr. The two figures
show that stable and unstable systems can be found all over the
diagram, but with a strong preference for unstable systems to have a
dominant binary, while stable unequal systems have a slight preference
for a dominant single. At young ages hierarchical triple systems
therefore frequently have dominant binaries. (c) A system with outer
period of 2 Myr will for the first time reach apastron after 1 Myr. In
the figure all systems in the grey shaded area have reached apastron
at least once within 1 Myr. During that time, no system has reached a
separation of more than 50,000 AU. The dotted blue line shows how far
the center-of-mass of a triple system has moved in a given amount of
time assuming a velocity of 1 km/sec.  Values are shown for 1 Myr and
10 Myr.  The widest systems, which take tens or hundreds of millions
of years to unfold, will have moved away from the denser and more
perilous environment in which they were born before being fully
unfolded.


\onecolumn

\clearpage

\begin{figure}[h]\begin{center}\leavevmode
\psfig{file=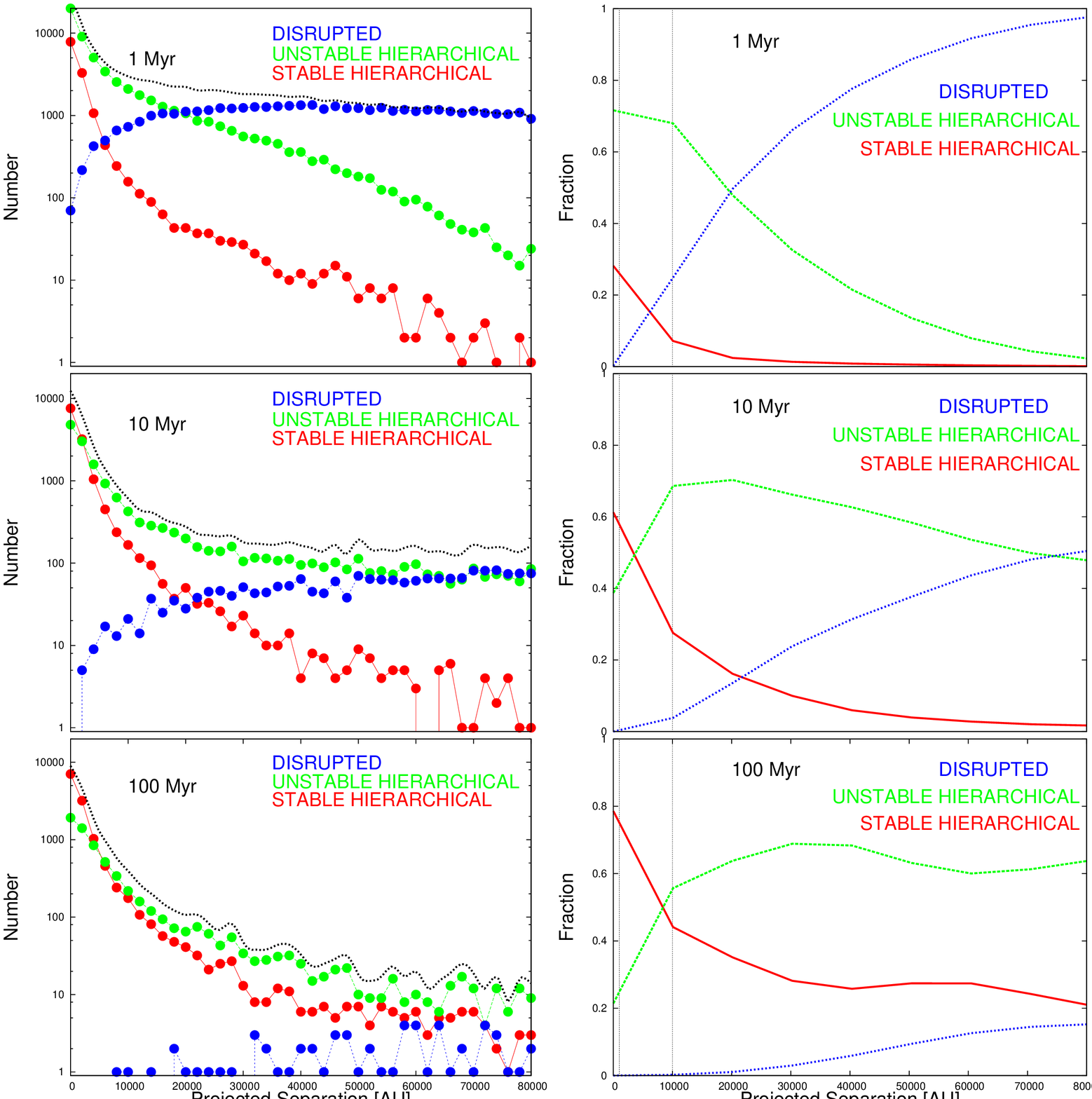,height=12cm}
\end{center}
\end{figure}

{\bf Figure 4. $|$ Frequency of stable, unstable, and disrupted triple systems as function of projected separation.}
{\em Left column}: The number of bound stable (red), bound unstable 
(green) and unbound (blue) triple systems as a function of projected 
separation for three different ages. The black dotted line indicates the 
sum of all three. This is an actual snapshot at 1, 10, and 100~Myr
of the random locations in their orbits of the distant third stars relative 
to the center-of-mass of the binaries projected onto the sky for all 
180,218 simulations.
{\em Right column}: The fraction of bound stable (red), bound unstable (green),
and unbound (blue) triple systems relative to the total number of triple 
systems as a function of projected separation for three different ages.
These diagrams allow a statistical determination of the state of an observed
triple system. At 1 Myr the bound systems dominate but only out to 
separations of 0.1~pc, after which the disrupted pairs strongly dominate. 
At these early ages, the
unstable triples are more common than stable triples by factors of
three or more. As time passes, the unstable triples break up and for
systems with observed separations of less than 4,000~AU the stable
triples slightly outnumber the unstable ones at 10~Myr. At the same
time, the disrupted systems (which will appear as common proper motion pairs) 
are moving apart and become dominant only at separations exceeding 4~pc.
Finally, at 100~Myr, the stable triples dominate out to separations of
about 8000~AU, but for larger separations the very widest binaries are
mostly bound but unstable systems, while the disrupted triples play
only a minor role at these more advanced ages. The two vertical lines
mark separations of 1,000~AU and 10,000~AU.  
      






{\Large\bf Supplementary Information}

\vspace{0.5cm}

{\large\bf 1. Code and Assumptions}
    
\vspace{0.2cm}
     
{\bf Stellar Masses}: We use an initial mass function$^{31,32}$ with
minimum and maximum mass values of 0.02 to 2.0 M$_\odot$, between
which the masses were selected according to the Chabrier probability
density function.  Initially the selected three bodies have identical
masses, but they are allowed to accrete from surrounding gas (see
below), usually altering at least one or two of the masses.  The three
bodies were positioned randomly in non-hierarchical configurations,
not exceeding a ratio of 5:1, and scaled so their mean initial
separations matched a number randomly picked between 40 and 400~AU,
values guided by observations. Subsequently the center of mass of the
three-body system was moved to the origin which also was the center of
the gas cloud.  Finally, the initial three-dimensional velocity
vectors were randomly chosen for each body and re-scaled so that the
virial ratio was 0.5 at the beginning of the simulations.

\vspace{0.2cm}
     
{\bf Gas Cloud and Accretion}: The three bodies initially move within
a cloud core described as a Plummer sphere with the potential
$=-M/\sqrt{r^2+R^2}$.  The core radius $R$ is set to 7500~AU, as
suggested by observations$^{33}$. For each simulation a core mass $M$
is picked randomly between 1 and 10 M$_\odot$. The stellar bodies are
allowed to accrete according to the Bondi-Hoyle prescription. Twice
the accreted mass is subtracted from the cloud core in order to mimic
the effect of outflow activity from young stars. Finally, to simulate
the effect of the diffuse interstellar radiation field, the remaining
gas disappears linearly with time over a period of 440,000~yr, which
is the duration of the Class~I phase determined from Spitzer
data$^{34}$. We do not consider any angular momentum of the accreting
material, which could affect the orbits of the closer pairs of
binaries$^{35,36}$, but our main focus here is the behavior of the
third body. This simplified treatment of the gas dynamics is less
realistic than a full-scale hydrodynamic simulation, but is necessary
in order to perform the hundreds of thousands of simulations required
to study statistically the complex dynamical evolution of triple
systems.  Consequently, the present numerical simulations do not
properly represent the Class~0 phase of the star formation process,
when the bulk of the stellar masses is rapidly built up.  Rather,
these calculations represent the Class~I phase, when the newly formed
stars have reached almost their final masses.
    
\vspace{0.2cm}
   
{\bf Integration}: The motions of the three-body system were
integrated using the chain regularization method$^{37}$ that provides
good accuracy in dealing with the 1/r$^2$ character of the
gravitational force as required for a precise treatment of frequent
close encounters.  Accretion effects were taken into account after
every integration step according to the Bondi-Hoyle prescription.  We
assumed the gas speed to be zero and thus the accretion causes
friction in addition to increasing the star masses.  After the gas
cloud has vanished entirely, the slowdown method$^{38}$ was used to
speed up the computation.

\bigskip
  
{\large\bf 2. Dynamical Evolution of Newborn Triple Systems}

\vspace{0.2cm}
     
It has been known for a long time that systems of three bodies are
unstable if they are in a non-hierarchical configuration. Such systems
will always evolve dynamically into either a stable hierarchical
system, or one member will escape and leave behind a bound binary
system$^{20,22}$.  This highly chaotic behavior of multiple systems
has been extensively explored numerically in the context of young
stars$^{21,39-42}$. The breakup of a young multiple system will most
often occur during the protostellar stage$^{24}$, and as a consequence
some of the ejected members may not have gained enough mass to burn
hydrogen, thus providing one of the key pathways for the formation of
brown dwarfs$^{43}$.  Detailed N-body simulations of newborn triple
systems still embedded in their placental cloud cores show that
protostellar objects are often ejected with insufficient momentum to
climb out of the potential well of the cloud core and associated
binary. These loosely bound companions can travel out of their dense
cloud cores to distances of many thousands of AU before falling back
and eventually being ejected into escapes as the cloud cores gradually
disappear and the gravitational bonds weaken. Protostellar objects
that are dynamically ejected from their placental cloud cores, either
escaping or for a time being tenuously bound at large separations, are
dubbed {\em orphaned protostars} and offer an intriguing glimpse of
newborn stars that are normally hidden from view$^{23}$.  A number of
such orphans have been identified in nearby star forming regions in
the vicinity of deeply embedded protostars, for the first time
allowing detailed studies of protostars at near-infrared and even at
optical wavelengths.

The role of the cloud core is important. It is a rarely appreciated
fact that without an additional gravitational potential,
non-hierarchical triple systems will virtually always break apart into
a stable binary and a third member that escapes the
system$^{22,44}$. In order to form a hierarchical
system where all three members are bound, an additional potential is
needed.  This can be provided either by the nascent cloud core, or by
additional stellar bodies, such as in quadruple or higher-order
multiple systems.

For comparison with observations, it is important to recall that the
higher the eccentricity is, the longer will the third body stay at
distances larger than the semimajor axis of the orbit. 

In recent years much discussion has centered on the ejection of
planets from forming planetary systems. We note that such processes
are dynamically very different from those discussed here. In a stellar
triple system the masses of the bodies are generally large and within
one or two orders of magnitude comparable. This generates strong and very fast
interactions, resulting in a rapid dissolution of the system. In
contrast, for a forming planetary system the planets have minuscule
masses compared to the central star and thus are in orbit around the
star. Their orbital evolution is secular, only gradually changing
until the planets approach orbital resonances, at which time an
ejection under the right circumstances may become possible.

\vspace{0.2cm}

\bigskip
  
{\large\bf 3. Classification of Binaries}

\vspace{0.2cm}
     
There is a rather well established classification of binaries
depending on observable or physical characteristics (e.g. visual
binaries, eclipsing binaries, spectroscopic binaries, low-mass X-ray
binaries, W~UMa binaries, etc.). In contrast there is very little agreement
on the classification of binaries as a function of their separation,
even though this is perhaps the most important parameter for
determining the physical properties of a binary. We here list two
attempts of a classification based on separation:

Zinnecker proposed this nomenclature$^{45}$ ($P$ is the orbital period
in years):\\

Extremely close binaries:  $P$ $<$ 10$^{-3}$\\
\indent Very close binaries: 10$^{-3}$ $<$ $P$ $<$ 1\\
\indent Close binaries:  1 $<$ $P$ $<$ 10\\
\indent Wide binaries: 10 $<$ $P$ $<$ 10$^2$\\
\indent Very wide binaries: 10$^2$ $<$ $P$ $<$ 10$^3$\\
\indent Extremely wide binaries: $P$ $>$ 10$^3$\\

With steadily improving observational techniques it is now possible to
determine the (projected) separation of most binaries, and hence a
more practical classification scheme has been proposed by Goodwin$^{46}$
($a$ is the semimajor axis in AU):\\

Close binaries: $a$ $<$ 50 \\
\indent Intermediate binaries: 50 $<$ $a$ $<$ 1000\\
\indent Wide binaries: 1000 $<$ $a$ $<$ 10000\\
\indent Very wide binaries: $a$ $>$ 10000\\

This classification is particularly relevant for binaries formed in
dense clusters, where the stellar density is high and leads to
dynamical processing of the initial binary population$^{47}$.  In the
present paper we have adopted this nomenclature.

We note that observationally the very wide binaries correspond to
common proper motion pairs, and so it is extremely difficult to
determine if a given very wide binary is bound or disrupted, which
could raise a semantic question about what constitutes a true binary.

The lack of a widely accepted nomenclature has led to a rather arbitrary
use of the prefix ``wide'' in the literature.

\bigskip

{\large\bf  4. Are all Wide Binaries Triple Systems?}

\vspace{0.2cm}

This question can be addressed from an observational
  and a theoretical perspective. Observationally, wide binaries are
  often found to be triple systems$^{8,9,48}$. However, most wide
  binaries are not known to be triple systems$^{49}$. This might
  reflect an intrinsic property of wide binaries, but also reflects
  the fact that only for nearby wide binaries (up to $\sim$30~pc) does
  the combination of present-day direct imaging and radial velocity
  studies cover the full separation range for companions.  In other
  words, the detection of triple systems is severely
  incomplete$^{26}$. The question is not likely to be answered
  empirically in the foreseeable future.

Theoretically, there are two aspects. First, while
  wide binaries can naturally form via dynamical evolution of triple
  systems, this does not imply that other formation mechanisms do not
  operate (see the following section), and other mechanisms can form
  wide binaries that are not triple systems. Second, as pointed out
  earlier, the close binary will during the protostellar and pre-main
  sequence phases be surrounded by significant gas in the cloud core
  and by circumstellar gas, and the dynamical friction will lead to
  gas induced orbital decay$^{27,50}$.  If the binary becomes bound
  shortly after birth of the triple system then a merger of the binary
  components is possible. These events will not affect the third body
  that has been ejected into a distant orbit, and so the final result
  is a wide binary with only two stars.  

The answer to the question is therefore ``no''.

\bigskip
 
{\large\bf 5. Formation of Wide Binaries in Dissolving Clusters} 
    
\vspace{0.2cm}
     
Many binaries are likely formed through disk fragmentation, which
readily explains the existence of some close and intermediate
binaries.  It is also generally accepted that core fragmentation plays
a critical role in the formation of stars and binaries, but with
typical core sizes of several 10$^3$~AU, fragmentation fails to
explain systems larger than this. Independent star forming events in
cores with larger separations might conceivably lead to bound systems,
but even if this were to occur, the pair would likely not remain bound
as the gas disperses. Additionally, most stars form in clusters, where
stellar separations typically are less than a few thousand AU, and so
binaries approaching such sizes would promptly be destroyed through
dynamical interactions. In short, the existence of binaries with very
wide separations poses a challenge to models of star formation.

Recently an interesting theory for the formation of very wide binaries
has been proposed$^{5}$. As gas is dispersed in a newborn cluster of
stars, the cluster will rapidly expand, leading to loss of stars.  Two
initially unbound stars (or, for that matter, a star and a binary) may
find themselves closely associated in phase space and thus form a
binary. The upper separation limit is set by the size of the cluster,
which is of the order of 0.1~pc. As the cluster potential becomes less
important, the pair may remain bound as it drifts away. The resulting
wide binary fraction is very sensitive to the initial conditions.
In a similar study, the long-term survival of the
  wide binaries was examined$^{6}$.  At any given time, a cluster will
contain a transient population of weakly bound pairs, which are
perturbed into and out of formally bound states.  To evaporate intact
from the cluster, a pair must form in the outskirts of the cluster.
The total number of such wide binaries is not sensitive to the cluster
population, and is about 1 pair per cluster.  The dissolution of many
small clusters of typically a few hundred stars is therefore likely to
contribute more very wide binaries than larger but more rare
clusters.

It is likely that the cluster evaporation mechanism is
contributing to the field population of very wide binaries. We note that
the wide binaries formed this way may or may not be triple systems,
that they are not primordial, in the sense that the stars are born in
separate collapse events, and that the binaries are formed, i.e.
become bound, with their wide separations.

In marked contrast, for the triple mechanism espoused here, {\em (a)}
the very wide systems are primordial, i.e. all members are formed in
the same collapse event, and {\em (b)} the systems are formed in a
compact configuration and expand as they are drifting away, and for
the widest binaries may unfold to their largest dimensions only after
they have escaped from their dense and perilous nascent environment.

\bigskip
  
{\large\bf 6. Unfolding of Wide Binaries}
    
\vspace{0.2cm}
     
A key aspect of the triple decay mechanism is that all three bodies
in a wide or very wide system were born from the same collapse event,
in close proximity despite their current enormous separations.  Once
the dynamical interactions transforming the three bodies from a
chaotic non-hierarchical configuration to a stable hierarchical system
have taken place, then the orbital parameters are set, and the bodies
will follow well determined orbits. While the inner binary in most
cases will have a rather short period, the outer body can take very
long (more precisely half an orbital period) before the system for the
first time has fully unfolded.

Figure~3c shows the maximum extent $a(1+e)$ of the simulated triple
systems as a function of half the orbital period of the systems.  It
will take any system half the orbital period to reach its full extent
for the first time.  As an example, the vertical dashed line marks an
age of 1~million years. All systems in the grey area to the left of
the line will have reached their apastron at least once, and, for the
shorter period systems, many times. But none of the systems to the
right of the vertical line will have reached their first apastron
passage and are thus not fully unfolded at 1~Myr. Because stellar
masses are factors in Kepler's 3.~law, and because the maximum extent
also depends on the orbital eccentricity, the relation between maximum
extent and time has a non-negligible width. As the figure shows, all
systems with full extent of less than 13000~AU have fully unfolded
after 1~Myr, while no systems with full extent larger than 50000~AU
have fully unfolded. 

During the time that the system is unfolding for the first time, it
also drifts gently away from its birth site. Typical turbulent
velocities in star forming clouds are around 1~km/sec, and
hence this is the velocity dispersion of the stars and multiples born
from the cloud. The blue dashed line in Figure~3c indicates how far
the center-of-mass of a triple system has moved in a given amount of
time assuming a velocity of 1~km/sec. In 1~Myr a system will have
drifted 1~pc and in 10~Myr it has drifted 10~pc. This is important,
because the space density of stars is higher at the birth sites of the
triple systems than in the general field. So by the time the very
widest systems have finally unfolded, they have drifted away from the
more perilous environment of their birth, diminishing their risk of
premature disruption.

\bigskip
  
{\large\bf 7. The Destruction of Very Wide Binaries}
    
\vspace{0.2cm}
     
Wide binaries are 'soft', i.e. they have binding energies that are
much smaller than the mean of the local stellar velocity
distribution$^{51}$.  While 'hard' binaries are resilient to
encounters with other stars, the soft binaries are sensitive to
breakup partly from (rare) close encounters with stars, but also from
the cumulative effects of distant but much more numerous weak
encounters$^{52}$.  Additionally, very wide binaries that pass near or
through giant molecular clouds are subject to immediate
disruption$^{16,17}$.  Finally, for very wide binaries the Galactic
tidal field will also lead to eventual dissolution of the
binary$^{53}$.  Such perturbations primarily have an impact on very
wide binaries with separations larger than 0.1~pc and on timescales of
Gyr, so the distributions shown in Figure~4 for much younger ages
would not be significantly influenced by this dynamic erosion.

The above assumes that the wide binaries have survived encounters at
their birth sites. Most stars are formed in clusters, where the
stellar density is much higher than in the comparatively empty space
exemplified by the solar neighborhood. Many binary systems are
therefore expected to be disrupted shortly after birth$^{54}$,
as is indeed observed in the binary separation distribution function
in the Orion Nebula Cluster$^{55}$.

\bigskip
  
{\large\bf 8. The Separation Distribution Function}
    
\vspace{0.2cm}

Binaries have separations that span more than a factor 1 million, from
close, short-period spectroscopic binaries to systems more than 0.1 pc
wide$^{56,57}$. The separation distribution function describes the
frequency with which binaries populate the various separations. \"Opik
proposed that this distribution$^{58}$ follows $f(a) \propto 1/a$, in
other words it is flat in $log(a)$, whereas Kuiper found a log-normal
distribution$^{59}$, which was later supported by the work of
Duquennoy \& Mayor$^{56}$, who found a peak around 30~AU. For wider
binaries however, several studies offer evidence that
the separation distribution more looks like \"Opik's law$^{60,61}$, at
least out to separations where the destruction of the widest systems
becomes significant. For the {\em inner} binaries in triple systems it
has been shown that the application of the statistical theory of the
three-body break-up leads to \"Opik's law of binary
separations$^{62}$.

Our simulations are, for practical reasons, stopped after 100~Myr. We
are thus following a population of triple systems that were all formed
in a single ``burst''. Observationally, this is well matched at 1~Myr
when one observes a star forming region, or at 10~Myr when one
observes a moving group. However, at 100~Myr the triple systems
presumably will have mixed with the general Galactic field population.
If the widest of these field systems have been gradually destroyed, then in
principle a more correct comparison would be with a population of
triple systems that have been continuously created, destroyed, and
mixed$^{17}$.

\begin{figure}[h]\begin{center}\leavevmode
\psfig{file=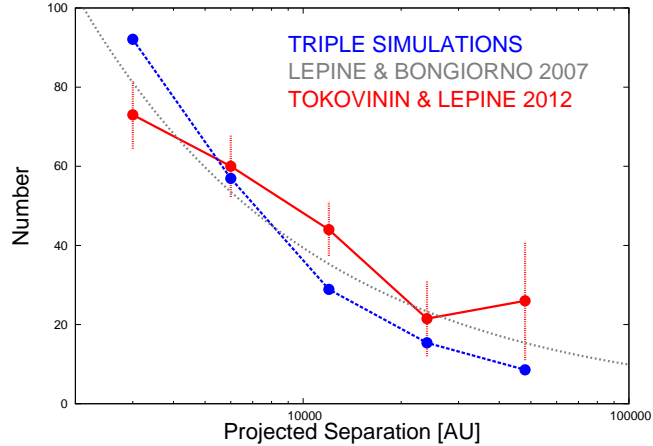,height=6.0cm}
  \caption{{\bf Figure S1. $|$ Comparison between simulations and
      observations.}  The red dots represent the most recent
    observations of wide binaries$^{15}$, the grey
    curve is a power-law$f(s)ds \sim s^{-1.6}ds$ fit to another data
    set of observations$^{14}$, and the blue dots are the triple
    simulations at 100 Myr sampled in the same bins as the red dots.
    The error bars on the observations represent sqrt(N) without
    including subtle systematic biases that are difficult to quantify.
    The errors on the simulations are smaller than the blue circles.
    The simulations include stable and unstable hierarchical systems
    as well as disrupted triple systems with projected separations in
    the selected separation ranges.}
\end{center}
\end{figure}

L\'epine \& Bongiorno studied wide common proper motion pairs from the
$Hipparcos$ catalog$^{14}$, and found that the projected separations
$s$ follow \"Opik's law $f(s)ds \sim s^{-1}ds$ only up to
3000--4000~AU, beyond which it falls more steeply, like $f(s)ds \sim
s^{-l}ds$ with $l$ = 1.6$\pm$0.1 out to $s \sim$ 100000~AU. Chanam\'e
\& Gould fit the distribution of wide binaries in the disk and
in the halo$^{13}$, and found a power law with $l \sim 1.67\pm0.07$ for the
disk and $l \sim 1.55\pm0.10$ for the halo, in excellent agreement
with the results of L\'epine \& Bongiorno$^{14}$.

In a recent study Tokovinin \& Lepine have determined the number of
wide binaries in five bins between 2000 and 64000~AU$^{15}$. We have
counted all triple systems with projected separations in these bins
and plotted them in Figure~S1 together with the data of Tokovinin \&
Lepine$^{15}$ and the power law $f(s)ds \sim s^{-l}ds$ with $l$ = 1.6
found by Lepine \& Bongiorno$^{14}$. The overall correspondence is
good, especially when considering that the observational data are
still incomplete for low-mass companions (which the simulations find
are abundant) and that the data may still be affected by systematic
biases.

\bigskip

{\large\bf 9. Compact Triple Systems}
    
\vspace{0.2cm} 

Triple systems are found with a huge range in separations between the
inner binary and the outer third body.  Our simulations have
demonstrated that the very widest binaries observed can be understood
as extremely hierarchical triple systems. But the question naturally
arises how very compact triple systems are formed.  Examples of
compact triple systems abound, one case is LHS~1070, where three mid-
to late-M dwarfs have separations of 3--9~AU$^{63}$.
The smallest mean separations in our simulations are of the order of
40~AU, and so do not naturally account for such compact systems. The
simulations, however, do not take into account that newly formed
binaries in dense cloud cores are surrounded by massive disks and
envelopes, so viscous interactions will sometimes shrink newly formed
binaries$^{64}$. This effect is sensitively
dependent on how early the close binary is formed; the later the two
components become bound, the less circumstellar material will be
present, and the less will be the effect of viscous interactions
(see also Section~4).

Additionally, a triple system can be the decay product of a quadruple
or higher-order system. Each time a body is ejected from a
non-hierarchical system the mean separation of the remaining bodies
grows smaller. This effect can be particularly important in a cluster
of stars, where small N-body systems initially are common. The
end-product of such higher-order decay can therefore be a very hard
triple system. The ejected stars may have been stripped of much of
their circumstellar material in the dynamical interactions$^{65}$, and
this could naturally account for the existence of diskless
stars$^{66}$.

\bigskip

{\large\bf 10. Examples of Wide and Very Wide Binaries}

\bigskip
  
{\large\em 10.1 Proxima Centauri}

The nearest known star to Earth is Proxima Centauri, discovered by
Innes$^{67}$, who noted the similarity of its high proper motion with
that of $\alpha$ Centauri, located at an angular distance of 2.2$^o$.
$\alpha$~Cen is a G2V star, and has a close companion with an orbital
period of 80~yr, while Proxima~Cen has a spectral type of M5.5V.  The
similarity of the distance to $\alpha$~Cen~A/B (1.33~pc) and to
Proxima~Cen (1.30~pc) and their similar motions have long suggested
that they may form a gravitationally bound system$^{68,69}$.  Using
the best available data, including from {\em Hipparcos}, it has been
concluded that the triple system is indeed likely to be physically
bound$^{4}$, with a physical separation currently of
15000$\pm$700~AU. In our model, despite their current very large
separation, the three $\alpha$~Cen components were born together in
the same collapse event, initially forming an unstable
non-hierarchical configuration. Shortly after birth, Proxima was
ejected into its current distant orbit.  At an age$^{70}$ of
$\sim$5.4~Gyr, evidently the $\alpha$~Cen triple system has achieved a
highly stable orbit. Even if the eccentricity of Proxima were as high
as 0.9, this would imply a periastron distance of $\sim$1500~AU, much
higher than the semimajor axis of the AB pair of about 22~AU, thus
ensuring that the system has remained stable.

\bigskip

{\large\em 10.2 Wide Binaries in Star Forming Regions}

Wide binaries have been known in star forming regions for a long
time$^{71}$, examples include Haro 1-14 (projected separation 1700 AU,
with the companion being a spectroscopic binary$^{72,73}$), the
non-hierarchical triple systems Sz~41 (320+1840~AU$^{74}$) and
LkH$\alpha$336 (2320+4320~AU$^{72,75}$), and the hierarchical triple
system SR~12 (1100~AU, where the distant component is a brown
dwarf$^{76}$).  Detailed studies of embedded protostars have revealed
a population of distant ($<$4500~AU) companions that become less and
less frequent as the protostars age$^{77-79}$, that is, many of these
components are lost already during the protostellar stage$^{23}$.  As
detailed imaging studies become more common, it is expected that more
such wide triple systems will continue to be found.

\clearpage

{\large\em 10.3 Wide Binaries in Young Moving Groups}

If a star forming region is not very massive, all its stars will
disperse after the original molecular cloud has disappeared, without
leaving a cluster behind. For a while, these young stars can be
identified as a moving group, where the members share kinematical and
physical properties, and all have the same age, typically 10-20 Myr.
Because members of young moving groups have not been part of a massive
cluster with violent dynamical interactions, binaries have a better
chance of surviving the time immediately after birth. In recent years,
careful analysis has led to the discovery of an increasing number of
wide and very wide binaries in moving groups, including 51~Eri$^{80}$,
TW~Hya$^{28}$, TWA~30$^{81}$, V4046~Sgr$^{82}$, T~Cha$^{29}$, and
RX~J0942$^{83}$.

This is consistent with the results presented here: after 10~Myr all
but the most extreme wide binaries have unfolded and the members of
the moving group have spread apart, typically having traveled
$\sim$10~pc from their sites of birth, and so it is much easier to
identify those stars that belong together in physical systems. Since
highly eccentric systems, once unfolded, spend most of their time near
apastron, and since many of the bound but unstable systems have not
yet disintegrated, there should consequently be a larger number of
wide and very wide systems at an age of 10-20 Myr than at any other
time.  We therefore expect that many more wide binaries will be
discovered in moving groups in the coming years. For a given projected
separation, Figure~4 allows to determine the probability that a given
wide system is bound and stable, or bound and unstable, or disrupted.

\bigskip

{\large\em 10.4 Wide Binaries in the Field}

The simulations presented here show that a non-negligible fraction of
the initial triple systems remain stable at an age of 100 Myr, when
the simulations are stopped. Most of the bound but unstable triples
have decayed by then, so the majority of triple systems that remain
are stable, and if they have not already broken up, the main threat to
their existence is likely to come from external perturbations rather
than internal instability. An example is the triple system HD~212168
where the third component is at a projected distance of
6090~AU$^{84}$. Another example is the $\alpha$~Cen/Proxima~Cen system
discussed in Section~10.1.

\bigskip
  
{\large\bf 11. Summarizing the Observational Predictions}

\vspace{0.2cm}

We here summarize the observational consequences of the triple decay
mechanism discussed in this paper:

[1] Although all very wide binaries formed through the
  triple decay mechanism have originated as triple systems, they are
  not necessarily all triples after completing their pre-main sequence
  evolution, because dynamical friction in a gas reservoir will cause
  orbital decay that in some cases can lead to mergers.  

[2] Most very wide binaries are likely to be found in young moving
groups because in star forming regions they have not all had time to
fully unfold, and in the field they are eventually destroyed, so the
peak is likely to be somewhere between 10 and 100 Myr.

[3] Many distant companions will be very low-mass
  objects, since they are preferentially ejected in three-body
  dynamics. However, the companions can also be binaries because a
  frequent outcome of dynamical triple interactions with accretion is
  the formation of a single more massive star and a lower-mass binary.

[4] The components of very low mass wide triple
  systems tend to have comparable masses (Figure~3a,b).  Wide {\em
    unstable} triple systems often have either components with
  comparable masses or a more massive binary and a distant low-mass
  single star.  Since unstable systems are more common than stable
  systems at ages of 1-10 Myr (see Figure~4), it follows that -
  especially in star forming regions - we should often find triple
  systems with a more massive binary and a distant single low-mass
  star.  Wide {\em stable} triple systems, in contrast, are much more
  uniformly distributed across Figure~3, perhaps with a slight
  preference for systems with unequal components to have one higher
  mass component associated with a distant lower-mass binary.  As the
  unstable triple systems eventually break part, we expect that the
  population of triple systems at young ages differ markedly from the
  old field population of triple systems.


\bigskip
    
\bigskip



\end{document}